\documentclass[11pt]{article}
\usepackage[latin1]{inputenc}
\usepackage[english]{babel}
\usepackage[namelimits]{amsmath}
\usepackage{amssymb}
\usepackage{amsmath}
\usepackage{amsthm}

\begin{document}
\title{New holographic dark energy and the modified Bekenstein-Hawking entropy}
\author{
Stefano Viaggiu,\\
Dipartimento di Matematica,
Universit\`a di Roma ``Tor Vergata'',\\
Via della Ricerca Scientifica, 1, I-00133 Roma, Italy.\\
E-mail: {\tt viaggiu@axp.mat.uniroma2.it}}
\date{\today}\maketitle
\begin{abstract}
We show that in the standard derivation of the holographic dark energy  some 
conceptual issues are present.
In particular, the formula used in the literature to avoid black holes is not suitable in expanding universes. 
A more suitable expression for a holographic motivated dark energy must contain the energy density of the remaining
matter content of the universe.
However, 
we show that under some reasonable hypothesis, we can obtain a new physically motivated
expression for the holographic dark energy. By considering an appropriate 
time-dependence of the saturation-level parameter, a de Sitter
phase arises.
Moreover, by adopting an argument similar to the original Bekenstein one,
our approach justifies a correction of the Bekenstein-Hawking entropy for non-static isotropic expanding universes.
Finally, we write down the equation of state of a black hole embedded in Friedmann spacetimes.
\end{abstract}

\section{Introduction}
A lot of observations during the past decade (see \cite{1,q,2}) strongly support
a present day accelerating universe. In the standard 
$\Lambda$CDM cosmological model, this acceleration is caused by the
the presence of the so-called dark energy expressed in terms of a
cosmological constant $\Lambda$ representing about $70\%$ of the present universe matter-energy. However, the physical origin of this constant is still obscure. 
Moreover, the reasons why $\Lambda$ is so small and begins to dominate only recently still remains mysterious. 

Many attempts are present in the literature \cite{3,4,5,6,7}, based on
quintessence models, k-essence models, phantom models to cite some of them, to alleviate the issues of the standard cosmological paradigm. Perhaps
the more intriguing alternative is given by the hologhraphic pronciple 
(see for example \cite{8,9,10,11}). By following this principle, one can obtain an expression
for the dark energy density depending on some unspecified size of the universe that can justify the smallness of the cosmological constant.
Thanks to its relevance in the field of dark energy
models, we analyze critically the standard derivation 
leading to the upper holographic limit for the energy density. In this context, the aim of this paper is to show the weakness
of the standard derivation together with a possible solution to alleviate these issues and its physical consequences.
In section 2 we analyze the standard derivation
together with our criticisms. In section 3 we attempt to amend the standard derivation by using exact theorems of general relativity
concerning black hole formation. In section 4 we study a modification of the Bekenstein-Hawking  entropy for  black holes embedded in
Friedmann spacetimes. 
Finally, section 5 collects some conclusions.

\section{Criticisms of the standard derivation}
In \cite{12} has been suggested that the limit set by the black hole formation can be applied in quantum field theory to justify a short 
distance cut-off. This idea has been analyzed in the context of dark energy models. In particular, denoting with
${\rho}_{\Lambda}$ the quantum zero-point energy density, the total amount of mass-energy in a region of size $L$
cannot exceed the mass of a black hole of the same size:
\begin{equation}
{\rho}_{\Lambda}=k\frac{3c^2}{8\pi G L^2},
\label{1}
\end{equation}
where the real constant $k\in(0,1]$ denotes the saturation level: for $k>1$ black hole arises.
This crude estimation presents several issues. 

First of all, the derivation of (\ref{1}) makes use of the well known condition for black holes 
suitable for static spherically symmetric spacetimes. In a static context, $L$ represents a radial radius in spherical coordinates, 
not a proper distance, while in the holographic context $L$ is a proper distance. In this regard,
some justification should be done for the fact that 
condition (\ref{1}) works when proper distances are used. 

A second criticism is that condition (\ref{1}) holds for a spherical black hole. General relativity itself cannot account
for a maximal allowed value for the energy-density 
by using black hole physics only. This can be obtained when 
quantum mechanics arguments are taken into account \cite{TV,TV2}. Moreover, 
its value depends on the shape of the collapsing object. For a spherically symmetric body immersed 
in a minkowskian spacetime, the bound (\ref{1}) is correct. But for a generical shape, this value can change.
In any case, if we consider Friedmann universes, sphericity is an exact symmetry
for the spacetime and then  seems to be a natural choice. 

A third criticism concerns the interpretation of condition (\ref{1}). This represents the maximal mass-energy density
excess within a certain volume allowed to avoid a black hole. Thus we have a given unperturbed background (minkowskian for
(\ref{1})) and a perturbation representing the collapsing matter. The Schwarzschild solution represents  an exact spherically symmetric
perturbation of minkowskian spacetime. Hence, when we apply the condition (\ref{1}) to Friedmann universes, the second member of
(\ref{1}) must represent the energy density excess of finite extension
with respect to a given unperturbed configuration, otherwise the naive use of the formula 
(\ref{1}) is not suitable and can also lead to wrong results.  
To this purpose, consider a Friedmann flat background filled with ordinary dust dark matter
${\rho}_m$ and a spherical perturbation given by a time constant energy density, i.e. 
${\rho}_{\Lambda}=kc^2\Lambda/(8\pi G)$. Condition (\ref{1}) for the holographic cosmological constant it gives:
\begin{equation}
\Lambda =k\frac{3}{L^2}.
\label{2}
\end{equation}
Since the cosmological constant fills the whole universe, then one can in principle perform the limit 
$L\rightarrow\infty$ and then obtain $\Lambda=0$ in order to avoid black holes. But, as well known, in the standard cosmological model 
the presence of a small but finite cosmological constant does not lead to black hole formation. As a consequence, we cannot 
take, for example, the standard cosmological concordance $\Lambda$CDM model and then apply merely condition (\ref{1}).
Condition (\ref{1}) must thus be interpreted as an energy density excess perturbation of finite extension with respect to a given
background, i.e. in  our case a cosmological background not a minkowskian one.

Finally, the black hole formation also depends on the dynamics of the collapsing matter. Hence, it is natural to suppose that in an expanding
non-static universe the Hubble flow plays an important role  for the formation of black holes. This is certainly true if we look for black holes 
formation in a Friedmann context. In fact, there exist exact theorems in general relativity \cite{13,14,15,17} 
giving sufficient conditions for the 
non formation of trapped surfaces for exact spherical perturbations. As well known from the Hawking theorems \cite{16}, the 
presence of trapped surfaces is related to the formation of the black hole singularity. In the next section, with the help of
these theorems, we alleviate the weakness depicted above.

\section{Holographic principle from suitable black holes  theorems}
To our purposes, it is sufficient to consider the theorem shown in \cite{14,17} for open Friedmann flat cosmologies.\\
Consider, in a flat Friedmann cosmology with a background energy-density ${\overline{\rho}}_m$,
a spherical surface of proper radius $L$ and proper area $A$ and
a perturbation of proper mass $\delta M>0$ within $A$. Suppose that the current matter perturbation $\delta J_{\mu}$
is vanishing on the boundary of $A$, if
\begin{equation}
\delta M\frac{G}{c^2}<\frac{L}{2}+A\sqrt{\frac{G\;{\overline{\rho}}_m}{6\pi c^2}},
\label{3}
\end{equation}
then $A$ is not trapped. Note that $L, A$ are respectively the proper length and the proper area with respect to 
the perturbed configuration, i.e. backreaction is taken into account. Moreover,
the upper bound for the mass-excess depends on the energy-density of the background on which the
perturbation acts.\\ 
In the following, we make the reasonable assumption (radial inhomogeneities are not considered)
that the mass-excess within $A$ is spatially constant, i.e.
$\delta M=\delta\rho(t)4/3\pi L^3$. Under this assumption, by following the same reasonings leading to 
(\ref{1}), from condition (\ref{3}) we obtain the new holographic energy given by:
\begin{equation} 
{\rho}_{\Lambda}=k\frac{c^2}{G}\left[\frac{3}{8\pi L^2}+\frac{3}{Lc}\sqrt{\frac{G\;{\overline{\rho}}_m}{6\pi}}\right].
\label{4}
\end{equation}
Note that the new holographic dark energy (\ref{4}) 
depends on the matter content of the background that is supposed, for our purposes, to be 
filled with dark matter. This is in agreement with the fact that theorems at our disposal relate the proper mass-excess to the matter
content just present in the universe. Hence, any physically reasonable dark energy expression motivated by 
the holographic principle must be a function of the background energy density. 
In practice, the maximum energy-density mass excess {\it must depend on the matter-energy present in the universe and on its dynamics}.
In a Friedmann context, it is natural that the Hubble flow $H$ makes more difficult the black holes formation and as a consequence an higher value for the energy-density excess is expected.
Otherwise, the resulting dark-energy expression is conceptually wrong in light of black hole formation theorems. In fact,
note that the term involving ${\overline{\rho}}_m$ in (\ref{4}) can as well be
of the same order of the first term and depends, thanks to the Friedmann equations, on the Hubble flow $H$. This confirms that, according to physical intuition, 
in the presence of an expanding universe it is more difficult to build black holes.
 
The expression (\ref{4}) can be put into Einstein equations together with the density
${\overline{\rho}}_m$ and then solve them  with a suitable expression for $L$ and $k$. To this purpose,
thanks to (\ref{4}), for the running cosmological constant we have:
\begin{equation} 
\Lambda(t)=k\left[\frac{3}{L^2}+\frac{24\pi}{cL}\sqrt{\frac{G\;{\overline{\rho}}_m}{6\pi}}\right].
\label{5}
\end{equation}
According to the reasonings of the section 2, we cannot naively take the limit $L\rightarrow\infty$ in (\ref{5}).
In fact, in a given gedanken localizing experiment, reasonable limits must be imposed to the size of the localizing object, i.e. causality cannot be violated.
To this purpose, expression (\ref{5}) can be written in terms of the density parameters:
\begin{equation} 
{\Omega}_{\Lambda}(t)=
k\left[\frac{c^2}{H^2L^2}+\frac{2c\sqrt{{\Omega}_m(t)}}{HL}\right],
\label{6}
\end{equation}
together with the Friedmann equation ${\Omega}_{m}(t)+{\Omega}_{\Lambda}(t)=1$.\\
After calculating expression (\ref{6}) at the present time $t_0$ and solving with respect to $L_0H_0$ we have:
\begin{equation}
L_0H_0=\frac{c}{{\Omega}_{\Lambda 0}}\left[k\sqrt{{\Omega}_{m0}}+\sqrt{k^2{\Omega}_{m0}+k{\Omega}_{\Lambda 0}}\right].
\label{7}
\end{equation}
From a physical point of view, a value for $k$ such that $k\simeq 1$ seems
most desirable, but obviously it depends on the physical motivation that inhibits the  saturation with  the maximum allowed value for $\delta\rho$.

Moreover, we expect that the size $L$ is of the order
of the particle horizon or less
and not of the future event horizon. This is because a reasonable limit for an ideal experiment on large scales 
involving black holes is dictated by causality. The future event horizon struggles with causality. 
These choices are in agreement with the idea that dark energy has an origin similar to the one leading to
Hawking or Unruh effects, where the presence of horizons play 
an important role.
For a numerical example we set $k=1$ and the concordance values ${\Omega}_{\Lambda 0}\simeq 0.68, {\Omega}_{m0}\simeq 0.32$.
As a result we obtain $L_0\sim 2.4 c/H_0$, being $c/H_0$ the Hubble radius. Remember that  for the particle horizon
of the $\Lambda$CDM model
we have $L_0\sim 3.3 c/H_0$. Note that without the fundamental term involving ${\overline{\rho}}_m$, i.e. by using practically the expression  (\ref{1}), we obtain, for $k=1$,  $L_0\sim 1.2 c/H_0$. 

A viable model for holographic dark energy must also contain a phase, after the recombination era, 
where dark energy is negligible with respect to dark matter. At this epoch, by denoting with $L$ the particle horizon, we have
$L\simeq 3c/H$. Moreover, by considering soon after recombination ${\Omega}_{m}\simeq 0.96$ and ${\Omega}_{\Lambda}\simeq 0.04$,
from (\ref{7}) we obtain
$k\simeq 0.065$. As a consequence, by setting the saturation-level parameter $k$ as a truly time constant near to the limit
$k\simeq 1$, it is not possible to obtain an early matter dominated era, which is obviously in contradiction with cosmological data.
This is a typical problem plaguing holographic models. The way in which in the literature one attempts to alleviate 
this problem is to consider 
ad hoc interactions between dark matter and dark energy. 
Our approach strongly suggests that 
a physically viable expression for ${\rho}_{\Lambda}$, motivated by holography,  must depend on the other kinds (and their dynamics)
of matter-energy density present in the universe in a form dictated by (\ref{4}).
To alleviate this problem, the most simple assumption is to consider a time-dependent
$k(t)$ reaching the saturation value only asymptotically. In particular, it may be supposed that the
emission mechanism to create 'holographic' dark energy  is similar to the one of a black body radiation emitted at some radius
of the order of (or less)
the particle horizon. In this regard, $k(t)$ can be seen as a kind of efficiency parameter: a low $k$ indicates an inefficient 
emission, while a value near to the saturation value indicates that the emission is efficient as the one of a perfect black body
radiation, i.e. dark energy is an ideal holographic radiation. In fact, it is physically reasonable that the efficiency of
the emission process of the holographic dark energy depends on the presence of other kinds of matter: the process becomes more and more efficient as ${\overline{\rho}}_m$ decreases.  If we take for $L$ instead of the particle horizon the Hubble one, then by setting again
${\Omega}_{\Lambda 0}\simeq 0.68, {\Omega}_{m0}\simeq 0.32$, we have $k_0\simeq 0.32$.
We stress again that a time-running $k$ is certainly more justified in our context with respect to the old one given by
(\ref{1}).
This idea is not exotic in cosmology. In fact, during the radiation dominated era, before nucleosynthesis,
the radiation was in a black body configuration, as firstly conjectured by Gamow.\\
To be more quantitative, we are tempted to write:
\begin{equation} 
{\rho}_{\Lambda}=k\frac{4{\sigma}_{\Lambda}}{c^3}T^4=
k\frac{c^2}{G}\left[\frac{3}{8\pi L^2}+\frac{3}{cL}\sqrt{\frac{G\;{\overline{\rho}}_m}{6\pi}}\right],
\label{SB}
\end{equation}
In any case, independently on the correctness of the formula (\ref{SB}), the interpretation of the parameter $k$ as a time-dependent
efficiency parameter remains a viable possibility.
In this view,
the saturation value $k=1$ would correspond to an 'ideal'' holographic emitter. 
 
Hence, 
if we consider $L$ of the order of
the particle horizon (or less, the Hubble radius), the variability of $k$ with time is essential to regain agreement with real astrophysical data.
In alternative, we can use for $L$ the expression of the future event horizon with a constant $k$
but, as stated above, it is plagued by causality 
problems that have not yet been addressed. 

Concerning the Hubble radius proposal,
as well known \cite{9}, by using the  Hubble radius for $L$ with the old expression (\ref{1}) and a constant $k$, we cannot obtain dark energy.
However, note that the arguments of \cite{9} do not apply in our new holographic paradigm because $k$ is a time function and hence
${\rho}_{\Lambda}$ is not merely proportional to $H^2$ and thus does not scale as a dust-like solution. Nevertheless,
when $L=c/H$,  the dark density  (\ref{SB}) assumes the expression
\begin{equation} 
{\Omega}_{\Lambda}(t)=
k(t)\left[1+2\sqrt{{\Omega}_m(t)}\right].
\label{HR}
\end{equation}
Since  ${\Omega}_{m}(t)+{\Omega}_{\Lambda}(t)=1$, from (\ref{HR}) we get:
\begin{equation} 
k(t)=\frac{(1-{\Omega}_m)}{(1+2\sqrt{{\Omega}_m})}.
\label{HR2}
\end{equation}
The expression (\ref{HR2}) has the right behaviour for an emissivity parameter: at the recombination 
$k\simeq 0$ and a late times $k\rightarrow 1$. At the present time $k_0\simeq {\Omega}_{m0}\simeq 0.32$, that could be an 
interesting coincidence.
Recently,  in \cite{Pad} has been shown that the Hubble radius has an unexpected link concerning the value of the dark energy,
although in apparently different context with respect to holography.
Note that this nice feature cannot hold for $L$ given by particle horizon, future event horizon or inverse square 
Ricci radius. Moreover, by using the Friedmann equations with the expression (\ref{HR}) we have:
\begin{equation}
{\overline{\rho}}_m=\frac{4H^2{(1-k)}^2}{G{\left[k\sqrt{\frac{32\pi}{3}}+
\sqrt{\frac{32\pi}{3}(k^2-k+1)}\right]}^2}.
\label{HR3}
\end{equation}
By inserting expression (\ref{HR3}) in (\ref{4}) we see that, for a running $k$,  ${\rho}_{\Lambda}$  cannot be a dust component. This happens only for $k$ constant. Hence our model  introduces a well defined link between the dark sectors of the universe
driven by the emissivity parameter. By solving equation (\ref{HR2}) for ${\Omega}_m(t)$
in terms of $k(t)$, we can obtain the empirical expression for $k(t)$ from the distance-redshift relation on the ligth cone.
This can be calculated also for the other cases, i.e. particle horizon, inverse Ricci square, agegraphic holography etc.
Note that the use of the Hubble radius is consistent with the inflationary period. Here 
${\Omega}_m=0$ and the Hubble radius is constant, i.e. $L=c/H_I$, where $H_I$ is the inflationary potential. In this context,
$k(t)<1$ in order to assures the Planck-level measured non-gaussianity.

Concerning the far future limit $t\rightarrow\infty$,  it is easy to see that,
under suitable conditions on $k(t)$, also with $L$ the Hubble radius,
we can have a late time de Sitter phase. To show this, we write the equation for ${\rho}_{\Lambda}$:
\begin{equation} 
{\rho}_{\Lambda ,t}+3H{\rho}_{\Lambda}(1+{\gamma}_{\Lambda})=0,
\label{asli}
\end{equation}
where ${\gamma}_{\Lambda}$ is the  equation of state parameter for the holographic dark energy. As an example,
we look for solutions such that 
for $t\rightarrow\infty$,  $H(t)$ approaches a constant value $H_s\sim c\sqrt{\Lambda/3}$.
To this purose,  thanks to (\ref{HR3}),  note that for $t\rightarrow\infty$, $\sqrt{{\overline{\rho}}_m}\sim H(1-k)$. As a result, 
since $k\rightarrow 1$,
we have that in this asymptotic limit $H\sqrt{{\overline{\rho}}_m}\sim H^2(1-k)=o(H^2)$. Under the conditions 
above, the first term in (\ref{4}) is dominant with respect to the one involving ${\rho}_m$. Finally, by putting the first term of 
(\ref{4}) in (\ref{asli}), the desired asymptotic behaviour for $k(t)$ such that
$H\rightarrow H_s$ and ${\gamma}_{\Lambda}\rightarrow -1$ is
\begin{equation} 
k(t)={\left(\frac{H_s}{H}\right)}^2+o(1).
\label{sol}
\end{equation}
We stress that the behaviour (\ref{sol}) for $k(t)$ must hold only asymptotically for $t\rightarrow\infty$. Nevertheless, note that also
by taking $k(t)\sim 1/H(t)^2$ after recombination, from expression (\ref{HR3}), we see that the dark-energy density
(\ref{4}) looks like a constant with added a monotonically decreasing term.
Summarizing, 
a time-dependent $k$ can account for an early and a later de Sitter phase together with a dark matter dominated era also by 
using the mistreated Hubble radius in the holographic context. 
 
In any case, we stress again that the crude estimation (\ref{1}), although can be justified in a more rigorous way, is not
suitable in a Friedmann cosmological context. As clearly shows expression (\ref{6}), the term involving 
${\overline{\rho}}_m$ is at least at 
present time, of the same order (but greater) of the old usual holographic term and therefore cannot be neglected. 

We leave in a further paper the study of the efficiency paramater in terms of the astrophysical data by using for $L$ the particle horizon
and Hubble radius expressions.

\section{Bekenstein-Hawking entropy in expanding universes}

As a first consideration of this section,
note that the reasoning leading to the expression (\ref{1}) can be formulated in terms of the well known Bekenstein-Hawking entropy,
initially proposed for black holes in a vacuum asymptotically flat spacetime.
Hence, a first interesting consequence of our result is that 
expression (\ref{4}) suggests a correction term caused by the degrees of freedom that are due to the 
non-static nature of Friedmann spacetimes. Denoting with $k_B$ the Boltzmann constant, with $L_P$ the Planck length and
with ${\rho}_m$ the matter-energy content of the universe and adopting the original
Bekenstein argument \cite{Bek}, we have:
\begin{equation}
S_{BH}=\frac{k_B A}{4 L_P^2}+\frac{k_BA^{\frac{3}{2}}}{cL_P^2}\sqrt{\frac{G{\rho}_m}{6}}.
\label{entr}
\end{equation}
The correction term takes into account
the degrees of freedom of a non-static (flat) Friedmann universe. By using theorems present in \cite{13,17},
the expression (\ref{entr}) can also be generalized to a Friedmann universe 
with negative curvature: in this case another positive term arises proportional to $\sim H^2A^2$. As well known, the identification
of the surface $A$ with the analogue of the event horizon in the Schwarzschild case
in expanding universes is not (see for example \cite{wit}) a simple task. However, the theorem (\ref{4}) leaves 
undeterminated the nature of the enclosing surface $A$: it can be as well a ligth-like surface that can be identified, for example, with the 
apparent horizon of the black hole. These issues, although important, play no role at this stage of our formulation of the 
black hole thermodynamics in expanding universes. 
  
In a static universe, the added term is vanishing. Moreover, if all the energy-densities present in the universe $\sum_i{\rho}_i$ 
satisfy the Friedmann equation 
$H^2=8/3\pi G\sum_i{\rho}_i$ (i.e. we have a spherical black hole embedded in a Friedmann expanding universe, see
\cite{wit, mac}), 
then expression (\ref{entr}) becomes
\begin{equation}
S_{BH}=\frac{k_B A}{4 L_P^2}\left(1+\frac{H}{c}\sqrt{\frac{A}{\pi}}\right).
\label{entr2}
\end{equation}
Note that, by taking for $A$ the Hubble area, the entropy (\ref{entr2}) becomes  3 times the usual 
expression.  According to holography, an higher admissible upper bound for the density implies an higher upper bound for the 
entropy.
Generally, the correction term is negligible when $A<<c^2/H^2$. When the dimensions of an object become comparable with
the Hubble radius, this correction cannot be neglected.
We expect that the added term will be relevant also at the early stage of the universe, near  the Planck era.

It is interesting to study the fate of (\ref{entr2}) near the big bang. The first term is always vanishing at $t=0$. Concerning the added term, 
if $H\sim 1/t$, then for a power law cosmologies with $a(t)\sim t^{\alpha}$, for $\alpha\in (0, 1/3)$, 
$S_{BH}\rightarrow\infty$. Conversely, for a spacetime  with $\alpha>1/3$,  
$S_{BH}\rightarrow 0$. Interestingly enough, for stiff matter, i.e. $\alpha=1/3$, then entropy reaches at the big bang a finite non vanishing limit. 

As a further remark, we analyze the expressions (\ref{entr}) and (\ref{entr2})  from the point of view of the first law of thermodynamics. 
To this purpose, note that
in (\ref{entr}) the term $A^{3/2}$ can be written also as $V$. Since we are in a spherically symmetric context,  in (\ref{entr})
we have firstly explicitally written this term as a function of the proper area.
However,  in ligth to first law of thermodynamics, could be more natural and useful to express the added term as a function of the 
proper volume of the apparent horizon of the black hole. With this choice, we have:
\begin{equation} 
S_{BH}=\frac{k_B A}{4 L_P^2}+\frac{3k_B}{2cL_P^2}V H.
\label{termo}
\end{equation}
By differentiating (\ref{termo}) we get
\begin{equation}  
dS_{BH}=\frac{k_B}{4L_P^2} dA+\frac{3k_B}{2cL_P^2} V dH+
\frac{3k_B}{2cL_P^2}H dV.
\label{termo2}
\end{equation}
The first two terms in the right side of (\ref{termo2}) can be interpreted as representing $1/T$ times
the internal energy of the black hole.
In particular, the one  involving $dH$ can be seen as the increases ($dH\geq 0$) of the internal energy 
due to the expansion of the universe caused by the presence of some unspecified kind of matter.
The term proportional to $dV$ can be seen as a work term due to the Hubble flow. In particular, we can write:
\begin{equation}  
\frac{P}{T} =\frac{3k_B H}{2c L_P^2},
\label{w1}
\end{equation}
where $P$ denotes the pressure. For a universe with negative curvature, a further positive term proportional to $H^2A^2$ also appears in 
(\ref{w1}). 
Unfortunately, 
we have not been able to obtain a convincing expression for the temperature $T$ from general considerations. In fact, also for  
an isotropic expanding universe, we have not 
at our disposal an universally accepted
analogous of the surface gravity parameter for a black hole in 
an asymptotically flat spacetime. 
To do this, an explicit black hole solution in expanding universes can be useful. To this purpose, we stress again that
all the 
distances, areas must be calculated from the exact solution representing a spherically symmetric black hole 
embedded in an expanding Friedmann universe.
However, by denoting with $R_H$ the Hubble radius,
we can write formula (\ref{w1}) in the following form:
\begin{equation}  
P R_h L_P^2=\frac{3}{2}k_B T.
\label{ig}
\end{equation}
For an ideal gas we have $PV=N k_B T$, where $N$ is the particles number. 
It is interesting that in the formula (\ref{ig}) explicitally appears the apparent horizon. 
After multiplying both members of
(\ref{ig}) for the proper volume $V$ of the apparent horizon of the black hole, we have:
\begin{equation}  
P V=\left(\frac{3V}{2R_H L_P^2}\right)
k_B T.   
\label{ig2}
\end{equation}
Suppose now to decompose the 
proper volume $V(t)$ of the horizon in $n(t)$ elemenatary spherical cells of fixed proper radius $L_P$, i.e.
$V=4/3\pi n(t)L_P^3$. Then expression (\ref{ig2}) becomes:
\begin{equation}  
PV=\left(2\pi\frac{L_P}{R_H}\right) n(t) K_B T.
\label{cb}
\end{equation} 
The expression (\ref{cb}) generalizes the usual equation of state
suitable for ideal gases in the context of black hole thermodynamics in Friedmann flat expanding spacetimes.  Formula 
(\ref{cb}) can have a nice physical interpretation. When the universe is cold, as at present time, many quantum degrees of freedom 
are frozen: this is described by the low ratio $L_P/R_H$ at present time.
But when the universe has been hot, more and more degrees of freedom have been 
excited ($L_P/R_H\sim 1$). 
At the Planck epoch, when $R_H\sim L_P$, we have $PV\sim n K_B T$, with $2\pi$ a geometric factor due to the 
sphericity of the black hole (there is not reason to put a sphere into a rectangular box). A similar phenomenon happens for the 
ordynary statistical mechanics. Moreover, note that $n(t)$ is an integer and so also $R_H/L_P=N_P(t)$, provided that
the Planck length is the minimum distance available in the real world. Hence the equation (\ref{cb}) can be written in the
expressive form:
\begin{equation}  
PV=\left(2\pi\frac{n}{N_P}\right)K_B T,
\label{trek}
\end{equation} 
a kind of Bohr-Sommerfield quantization rule for the black hole equation of state. We do not speculate further on this formula.

The presence of a volume term in (\ref{termo2}) could struggle with the holographic principle. However, note that
the work term $PdV$ of usual thermodynamics arises thanks to a volume dependence of the entropy, and is unavoidable.
Otherwise, no work would be associated to the expansion of $A$, that seems rather unlikely.

As a further remark, note that all the reasonings of this section are a consequence of the theorem leading to
(\ref{4}) and thus do not depend on the interpretations concerning the nature of the holographic dark energy.

As a final consideration,
it should also be noticed that the theorem (\ref{4}) remains valid if we substitute the energy-density
${\overline{\rho}}_m$ with a constant energy-density, i.e. in a de Sitter expanding universe. In such a case, we have
again the formula (\ref{termo}), but with $H(t)$ the constant de Sitter value
$H=\sqrt{\Lambda/3}$. In this case, the term involving $dH$ in (\ref{termo2}) is vanishing, and as a result the internal energy of a 
black hole in a de Sitter universe is left unchanged with respect to the asymptotically flat case. 

 \section{Conclusions}
In this paper, we have discussed the usual calculations supporting the holographic dark energy expression.  
The usual derivation is plagued by the use of expressions for black holes formation that are not
suitable in a cosmological context. In fact, 
espression (\ref{1}) used in the literature is suitable for the black holes formation caused by a spherically symmetric static 
configuration in the vacuum. Unfortunately, condition (\ref{1}) cannot work in a cosmological context. 
In an expanding universe, as well known from theorems of general relativity, the formation of trapped surfaces depends on the matter content of the universe and on its dynamics.
In particular, the Hubble flow $H$ plays an important role making black hole formation more difficult. In this context, by using 
the theorems present in \cite{14,17}, we can obtain a new expression for the holographic dark energy. In 
order to have a conceptually and mathematically correct expression for an holographic motivated dark energy,
it is essential its dependence on the background
density, for example dark matter after recombination.
A choice for $L$ based on the future event horizon is plagued by serious yet not resolved causality problems. 
Moreover, in these models the far future limit $t\rightarrow\infty$ is rather problematic and can lead to a spacetime singularity.

We have tested  our new holographic dark energy
by requiring that at present time the proper length $L$ be of the order of the particle horizon or the Hubble radius and 
that soon after recombination the dark matter dominated the dark energy. The only way to satisfy these requirements is to take
the parameter $k$ depending on the cosmic time $t$. In this regard, the parameter $k(t)$ can be seen as measuring the efficiency 
of the holographic dark emission. Also by using for $H$ the Hubble radius expression and for $k$ a suitable 
time-dependent function, we can account for an early and late times de Sitter phase together with a dark matter dominated era 
after recombination. Remember that in a flat Friedmann spacetime, the apparent horizon is nothing else that the Hubble radius.

It is important to note that
with our new view of the holographic dark energy, the time variability of $k$ can be more easily justified.

It is worth to note that also with the new holographic dark energy (\ref{4}), an interaction terms can be considered in the 
right hand side of equation (\ref{asli}) (and obviously also for ${\overline{\rho}}_m$).

As a consequence of our approach, we are in the position to generalize the Bekenstein-Hawking entropy in an expanding
universe. This allows us to write the equation of state of a black hole (equation (\ref{cb})) together with a possible physical motivation.
Further investigations are needed to a more deep physical understanding.

As a final hint of this paper, we can conclude with the claim that 
all holographic motivated reasonings in a cosmological context must take into account, from the onset, 
the dynamical nature of the spacetime due to the Hubble flow.

\end{document}